# Conservation of a spectral asymmetry invariant in optical fiber four-wave mixing

*Anastasiia Sheveleva, [1] Pierre Colman, [1] John M. Dudley,[2] and Christophe Finot [1,*]*

1. Laboratoire Interdisciplinaire Carnot de Bourgogne, UMR 6303 CNRS-Université de Bourgogne, Dijon, France

2. Université de Franche-Comté, Institut FEMTO-ST, CNRS UMR 6174, Besançon, France

**E-mail:** christophe.finot@u-bourgogne.fr

The conservation of spectral asymmetry is a fundamental feature of the ideal four-wave mixing process as it exists in a medium combining quadratic chromatic dispersion and third-order nonlinearity. We test in this paper the robustness of this invariant in an experimental configuration where the excitation conditions of an optical fiber are sequentially updated, mimicking infinite propagation. This theoretical and experimental study reveals the high sensitivity of the asymmetry to very slight deviations from the ideal case, and we show that our idealized system behaves as an intermediate case between the ideal case of non-cascaded four-wave mixing and propagation in a system governed by the nonlinear Schrödinger equation.

# 1. Introduction

Optical fibers, owing to their minimal losses and their extended interaction lengths, have emerged as a crucial tool for exploring the intricate nonlinear dynamics arising from the interplay of dispersion and nonlinearity.[1] A well-established model to describe light evolution in single-mode fiber waveguides is the nonlinear Schrödinger equation (NLSE), one of the seminal equations of physics that is applicable not only to fiber systems but also to ocean waves, plasmas and Bose-Einstein condensates to cite only a few examples.[2] Among the well-known stable solutions of the NLSE is the bright soliton, a coherent structure that celebrates in 2023 its 50[th] anniversary in fibers, and that propagates while preserving its temporal and spectral intensity profiles.[3] However, a range of other known NLSE solutions also exist, including the well-known family of solitons on a finite background, where the interaction with a continuous background results in periodic temporal or spatial localization,[4] or even double localization such as observed for the Peregrine soliton.[5] Coherent solutions on periodically modulated backgrounds, such as cnoidal or dnoidal waves, are also part of this rich landscape.[6]

The analysis of the NLSE in the frequency domain offers remarkable insights into the spectral properties of these coherent waves, and specifically the dynamics of the four-wave mixing process (FWM) that rules the periodic energy exchanges among evolving spectral components. In the context of a focusing nonlinearity, FWM leads to modulation instability [7] and the generation of high-repetition rate trains of ultrashort localized pulses.[8] In the simplest configuration, called degenerate four-wave mixing, it only involves a high-intensity pump wave interacting with two other wave components at frequencies symmetrically located on both sides of the pump. Neglecting the existence of higher-order interactions, this dynamics is ruled by a system of three coupled differential equations. It exhibits a recurrent behavior which is an example of the celebrated Fermi-Pasta-Ulam-Tsingou recurrence.[9] Although some analytical solutions exist,[10, 11] it is also convenient to interpret the wave interactions in the phase plane through two canonical conjugate variables [12, 13] that form closed orbits, ensuring that trajectories never venture beyond the boundaries of their respective orbits. Such recurrent features have been recently been quantitatively measured in experiments using a novel setup involving iterated propagation over short segments of fibers where periodic filtering of higher-order spectral sidebands allows the framework of three ideal coupled equations to be retained, [14] and avoids higher-order effects such as third-order dispersion,[15] distributed losses,[16] Raman scattering [17] or accumulation of amplified spontaneous emission. [18] Experiments have seen excellent agreement between theory and experiment over several tens of kilometers of propagation. In addition, this system has allowed data collection for training of neural networks, [19] and a simple but efficient framework to control the phase space trajectory has been proposed and validated.[20]

In this contribution, we further study the energy exchange dynamics between the spectral components in this system. In contrast to most of the literature, however, we do not restrict ourselves to a signal with a perfectly symmetric initial optical spectrum, and we place a particular focus on how the spectral asymmetry evolves longitudinally. Whereas the asymmetry is a constant of the ideal degenerate FWM, we stress that when propagation is ruled by cascaded interactions, a completely different picture is observed. More surprisingly, in the setup developed in [14] where significant cascade is not possible, the evolution of the asymmetry is also not conserved. A theoretical perturbation analysis is developed to explain the physical origin of the fundamental process at the origin of this discrepancy, and this is confirmed with experimental measurements.

## 2. Principle and numerical simulations

### 2.1. Governing equations, normalization and conserved quantities

We first review the theoretical description of ideal FWM dynamics in the NLSE. In an ideal single mode and loss-free fiber, the evolution of a slowly-varying electric field envelope $\psi(z,t)$ is governed by the nonlinear Schrödinger equation: [1]

$$i\frac{\partial \psi}{\partial z} - \frac{\beta_2}{2}\frac{\partial^2 \psi}{\partial t^2} + \gamma |\psi|^2 \psi = 0, \quad (1)$$

with $z$ being the propagation distance and $t$ the time in a reference frame traveling at the group velocity. The group-velocity dispersion is $\beta_2$ and the nonlinear Kerr coefficient is $\gamma$. In the focusing regime of propagation ($\beta_2$ negative), we can rewrite the NLSE in normalized form:

$$i\frac{\partial A}{\partial \xi} + \frac{1}{2}\frac{\partial^2 A}{\partial \tau^2} + |A|^2 A = 0. \quad (2)$$

Here, normalized propagation and co-moving time variables $\xi$ and $\tau$ are linked to the dimensional quantities in fiber optics by $\xi = z/L_{NL}$ and $\tau = t/\sqrt{|\beta_2|L_{NL}}$. The characteristic length scale $L_{NL}$ is defined as $L_{NL} = (\gamma P_0)^{-1}$ with $P_0$ the initial power, which in our case corresponds to the total average power of the evolving field i.e. taking into account pump and any sideband components.[1] The normalized field $A(\xi,\tau)$ is related to its dimensional equivalent $\psi(z,t)$ by $A(\xi,\tau) = \psi(z,t)/\sqrt{P_0}$.

We discuss the fundamental wave mixing processes in the NLSE by considering the injection of a modulated pump wave $A_0$ with two sidebands at relative offset frequencies $\pm\Omega$:

$$\begin{aligned}A(\xi,\tau) &= A_0(\xi) \\ &+ A_{-1}(\xi)\exp(i\Omega\tau) + A_1(\xi)\exp(-i\Omega\tau)\end{aligned}. \quad (3)$$

The carrier frequency $\Omega_0$ is omitted and normalized offset frequency $\Omega$ is related to dimensional offset frequency $f_m$ in Hz by: $\Omega = 2\pi f_m \sqrt{|\beta_2|/\gamma P_0}$. In general, the injection of such a modulated signal in an optical fiber leads to the generation of multiple additional sidebands, but the FWM interaction can be idealized and truncated to describe only pump and first sideband energy exchange with distance. It leads to only three coupled equations that are well-known in the field of parametric amplification [21] :

$$\begin{cases} -i\dfrac{dA_0}{d\xi} = \left(|A_0|^2 + 2|A_{-1}|^2 + 2|A_1|^2\right)A_0 \\ \qquad\qquad + 2\,A_{-1}A_1 A_0^* \\ -i\dfrac{dA_{-1}}{d\xi} + \dfrac{1}{2}\Omega^2 A_{-1} = \left(|A_{-1}|^2 + 2|A_0|^2 + 2|A_1|^2\right)A_{-1} \\ \qquad\qquad + A_1^* A_0^2 \\ -i\dfrac{dA_1}{d\xi} + \dfrac{1}{2}\Omega^2 A_1 = \left(|A_1|^2 + 2|A_0|^2 + 2|A_{-1}|^2\right)A_1 \\ \qquad\qquad + A_{-1}^* A_0^2 \end{cases} \qquad (4)$$

It is convenient to separate the complex spectral amplitude $A_i(\xi)$ into its modulus $a_i(\xi)$ and phase $\varphi_i(\xi)$ : $A_i(\xi) = a_i(\xi)\exp(i\,\varphi_i(\xi))$. Contrary to most studies that have previously investigated this problem,[1] we do not limit ourselves to the case where the initial sidebands have symmetric amplitude with respect to the pump. As seen in **Figure 1**, panel (a), it is possible to describe the three-line asymmetric spectrum in terms of three variables $\eta$, $\phi$ and $\alpha$ that are physically related to the fraction of the total power in the central frequency component, the sideband-pump frequency component phase difference, and the asymmetry between the lateral sidebands of the spectrum :

$$\begin{cases} \eta = \dfrac{a_0^2}{a_0^2 + a_{-1}^2 + a_{+1}^2} \\ \phi = \varphi_1 + \varphi_{-1} - 2\varphi_0 \\ \alpha = a_{-1}^2 - a_{+1}^2 \end{cases} \qquad (5)$$

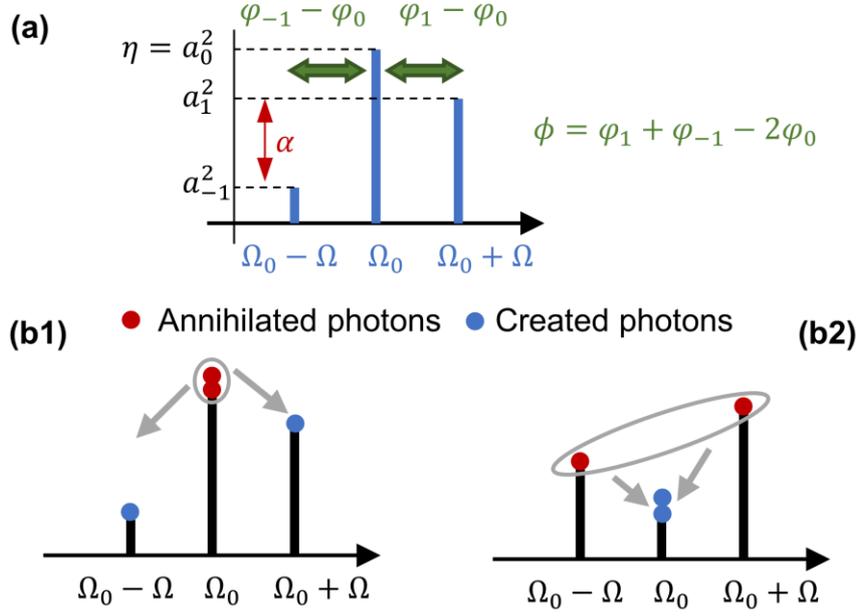

**Figure 1.** (a) Illustration of the parameters involved in the reduced degenerate model. (b) Illustration of the two possible processes of energy exchange between the three components within the frame of Equation (4).

In order to gain further insight into the nonlinear dynamics of this ideal system, it is essential to identify the invariants. First of all, given the conservation of energy in the lossless fiber, one may find a first invariant that is the average power, leading to:

$$a_1^2 + a_{-1}^2 + a_0^2 = 1. \qquad (6)$$

Combined with Equation (5), this leads to the normalized power of the sidebands that can be expressed as :

$$\begin{cases} a_0^2 = \eta \\ a_{\pm 1}^2 = \dfrac{1-\eta \mp \alpha}{2} \end{cases} \qquad (7)$$

Taking this relationship into account and given that $|\alpha|+\eta \leq 1$, the level of asymmetry that it will be possible to imprint for a given value of $\eta$ will be bounded.

A second crucial invariant of the system is the asymmetry $\alpha$, which arises from the Manley-Rowe relations. Indeed, as a continuous wave (CW) or modulation at $\pm\Omega$ are the only possibilities, the energy exchange among the waves has to be achieved through the two process shown in Figure 1(b). In one case, two photons from the pump generate one photon in each lateral sideband following the scheme $\Omega_0 + \Omega_0 \rightarrow (\Omega_0 + \Omega) + (\Omega_0 - \Omega)$ (Figure 1(b1)). In the other case (Figure 1(b2)), two photons at frequencies $\Omega_0 + \Omega$ and $\Omega_0 - \Omega$ lead to the emission of two new photons at $\Omega_0$. Therefore, the conservation of the power difference between the two sidebands is in both cases ensured.

A third conserved quantity is the Hamiltonian of the system. Indeed, it is possible to associate the problem with the one-dimensional conservative Hamiltonian [12]:

$$H = 2\eta\left[(1-\eta)^2 - \alpha^2\right]^{1/2} \cos\phi + (1-\kappa)\eta - \frac{3}{2}\eta^2, \qquad (8)$$

with $\kappa = -\Omega^2$ being a normalized mismatch and with the canonical conjugate variables being $\eta$, $\phi$ as defined in Equation (5).

## 2.2. The iterated segmented approach

The experimental observation of the canonical dynamics of ideal FWM as described by Equation (4) remains a difficult challenge. Indeed, the growth of higher-order sidebands through cascaded FWM leads to the formation of an extended spectral comb structure [22]. As a consequence, the ideal FWM process cannot be studied in isolation. To circumvent this limitation, inspired by [23, 24], we have proposed an iterated segmented approach [14] as shown in **Figure 2**. Initial conditions of three spectral lines, the phase and amplitude of which have been conveniently tailored, are launched into a short span of optical fiber (assumed in a first approximation as lossless). The fiber segment is sufficiently short such that additional sidebands located at $\pm 2\Omega$ cannot reach a significant level. After propagation in the fiber, the signal is spectrally filtered to exclude low-amplitude cascaded components that might nevertheless have developed. In order to compensate for possible losses induced by this filtering, the new three-component signal is reamplified before it can be reinjected within the fiber as new initial conditions.

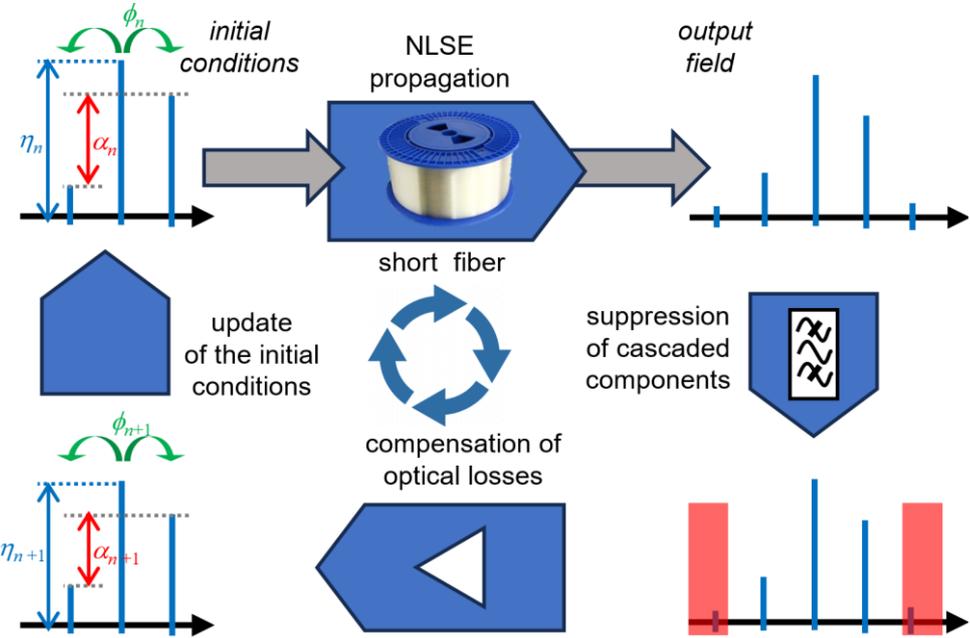

**Figure 2.** Principle of the segmented approach with iterated initial conditions.

## 3. Comparison of the evolution of the asymmetry in the various configurations

We now compare the non-linear dynamics that results from propagation governed by the full NLSE, the ideal FWM scheme and the iterated segmented approach described above. We devote special attention to the asymmetry parameter $\alpha$. In order to illustrate the propagation, we have selected the following parameters based on our experiments that will be discussed in section 4. We have considered an initial asymmetry level $\alpha_0$ of 0.15, and an average power leading to $\kappa = -1.2$. Note that for this value of $\kappa$, potential cascaded sidebands will not fall within the modulation instability gain bandwidth of the pump and therefore will not be efficiently amplified.[25] The fiber length involved in the segmented approach has a normalized length $\Delta \xi$ of 0.21. We considered two initial conditions associated with $\eta_0 = 0.75$, and phase $\phi_0$ of 0 or π. Plotting the dynamics in the cylindrical coordinate system ($\eta$, $\phi$, $\alpha$) fully captures the physics of this problem.

### 3.1. Results in the ideal FWM process

We first discuss the dynamics expected for the ideal case. **Figure 3** summarizes the evolution of the parameters observed for the two previously mentioned initial conditions located either to the right or to the left side of the separatrix separating bounded and unbounded phase behavior. Consistent with the conservation of the asymmetry parameter, the trajectory when represented in cylindrical coordinates is a fully closed orbit that is contained in the plane $\alpha = \alpha_0$, which is also confirmed by the panels (c) as well as by the panels (b) where the difference between the intensity of the lateral components located on either side of the pump remains identical with propagation.

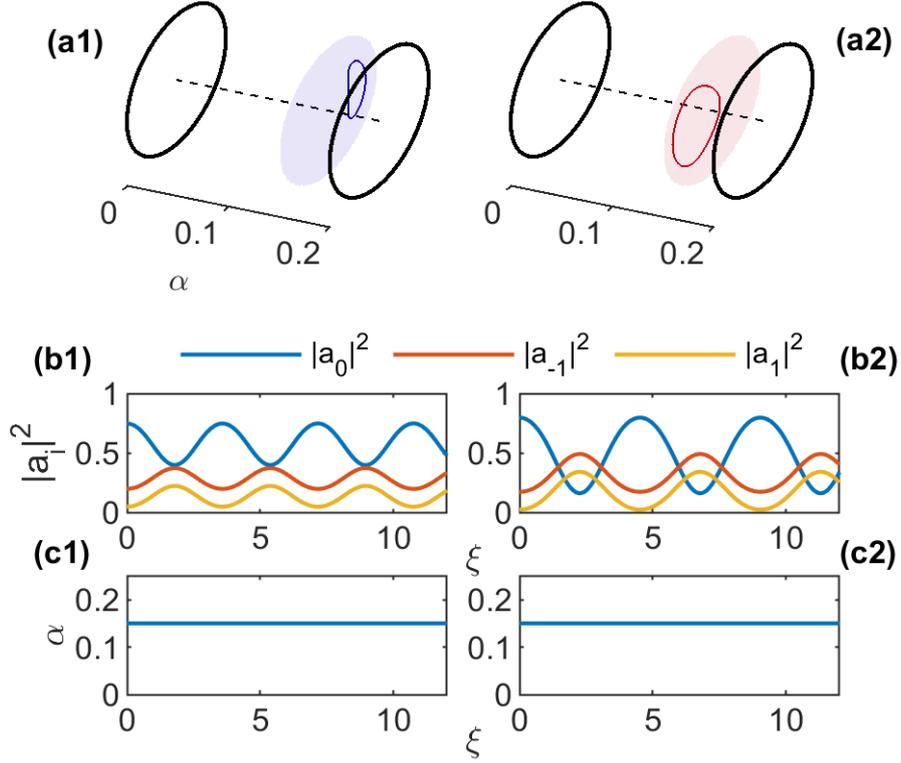

**Figure 3.** Results for ideal FWM (Equation (4)). Results obtained for an initial phase of 0 and $\pi$ are displayed on panels (**1**) and (**2**), respectively. (**a**) Trajectory plotted in the cylindrical coordinate system ($\eta$, $\phi$, $\alpha$). The plane corresponding to the initial conditions $\alpha_0$ is colored. (**b**) Evolution of the normalized intensity of the three spectral bands. (**c**) Evolution of the asymmetry parameter.

### 3.2. Evolution in the NLSE framework

Let us now discuss the evolution of the same parameters in the case where the dynamics are dictated by the NLSE and the FWM cascade gives rise to a significant number of new spectral components. The results obtained from the numerical integration of Equation (2) with the split-step Fourier algorithm are summarized in **Figure 4** and show striking differences from the ideal behavior discussed previously. First of all, it is clear that the asymmetry parameter as defined in the context of ideal mixing is in no way preserved. The trajectory is no longer a simple closed orbit, nor is it at all contained in a single plane. A much larger volume of the parameter space is now explored, and we can note in particular a change in the sign of the asymmetry parameter. This implies in particular a passage through a zero asymmetry value.

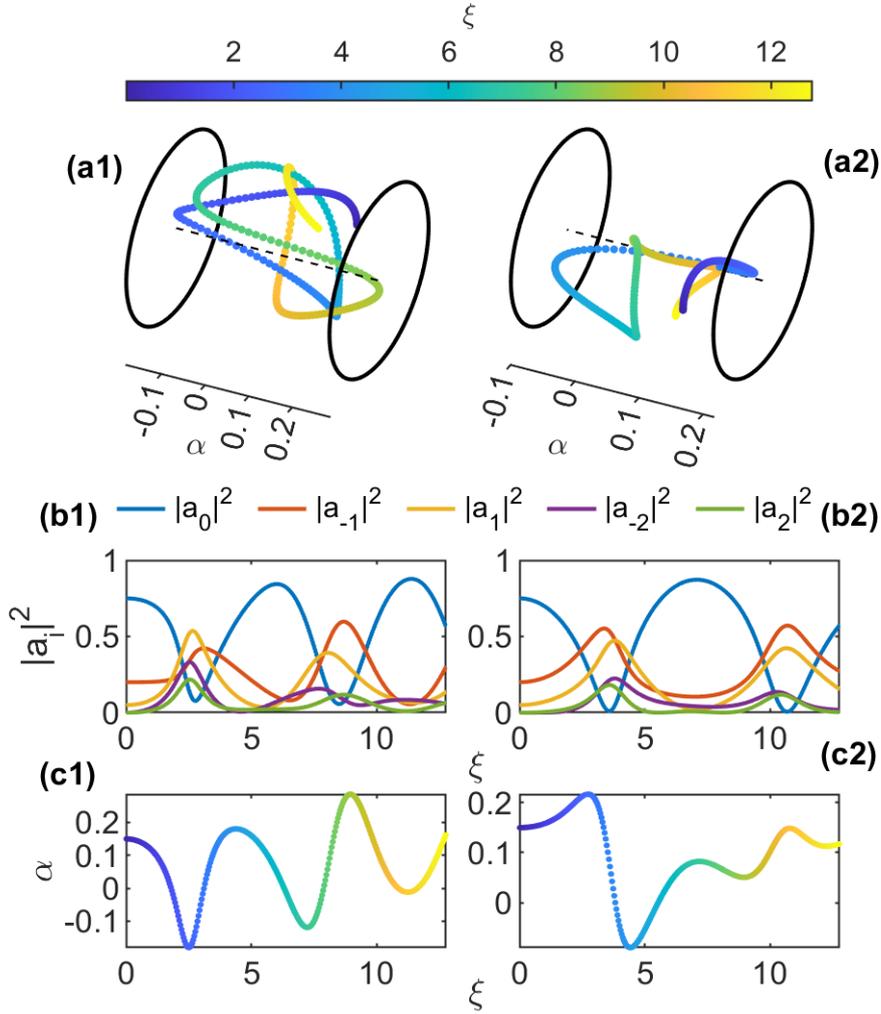

**Figure 4.** Results in the case of a propagation ruled by NLSE. Results obtained for an initial phase of 0 and $\pi$ are displayed on panels **(1)** and **(2)**, respectively. **(a)** Trajectory plotted in the cylindrical plot ($\eta$, $\phi$, $\alpha$). The distance of propagation is highlighted by changing the color of the symbols (see colormap). **(b)** Evolution of the normalized intensity of the five inner spectral bands. **(c)** Evolution of the asymmetry parameter (the colors refer to the distance as used in the colormap of (a)).

More details on the complete longitudinal evolution of the optical spectrum can be seen in **Figure 5**. In this case, there is a macroscopic built-up of energy in the $a_{\pm 2}$ and $a_{\pm 3}$ components (the level of the additional sidebands being normalized with respect to the level contained in the three central lines). Their levels are far from being negligeable : values up to 0.33 are reached for $a_{\pm 2}$ after propagation over a distance of 12.76 (30 km). The evolution of the extra-sidebands is complex and not necessarily synchronous. Panel (b) shows the spectrum at a distance of 1.79 (4.2 km) for which the asymmetry $\alpha$ is zero. We note that even though the first two lateral components have identical intensity, this is not the case for the higher-order components. Thus, the $a_{-2}$ component is almost twice its $a_{+2}$ counterpart. Consequently, the asymmetry has not disappeared, it has spread to the higher-order components. In other words, the simple metric based on $\alpha$ as defined by Equation (5) is no longer appropriate and an extended version is in this case relevant as discussed in [26].

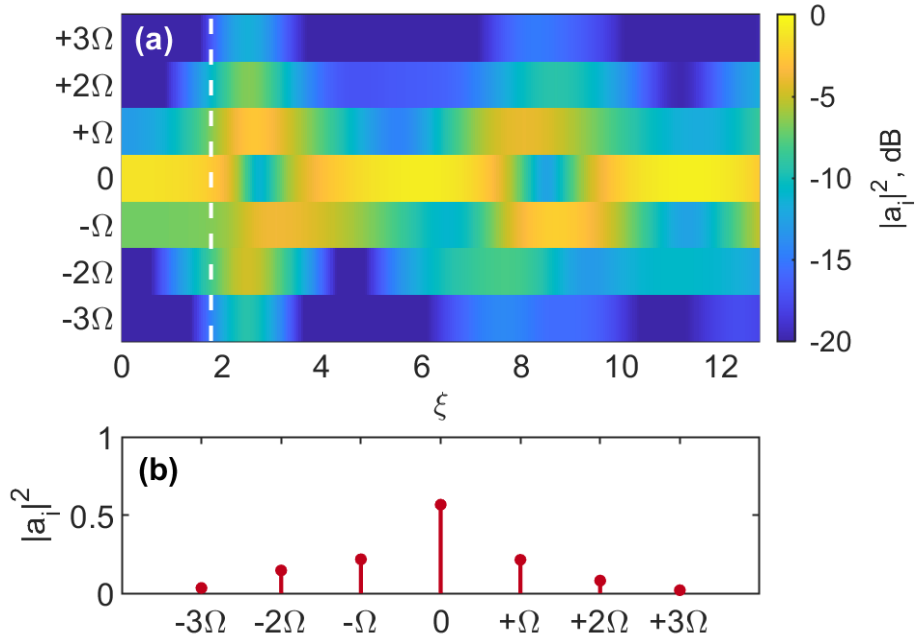

**Figure 5.** (a) Longitudinal evolution of the full spectrum for ($\eta_0 = 0.75$, $\phi_0 = 0$, $\alpha_0 = 0.15$). (b) Details of the spectrum when $\alpha = 0$ after a propagation distance $\xi = 1.79$ (see dashed line in panel (a)).

### 3.3. Evolution in the iterated segmented approach

By comparing the ideal model and the NLSE propagation, we observe to what extent the asymmetry of the first sidebands could have radically different evolution. In this part, we now discuss the evolution in the system introduced in section 2.2 where the growth of higher-order sidebands is suppressed at short and regular intervals, thus blocking the appearance of significant cascades. Results over a propagation distance of 12.76 corresponding to 60 segments of length $\Delta\xi = 0.21$ are shown in **Figure 6** and reveal an intermediate trend between the ideal case and the NLSE dynamics. Even though the asymmetry parameter is not preserved, we see that its evolution presents a much less complex structure than in the NLSE case, and the evolution of the powers in the different bands regains a simpler structure approaching the periodic behavior observed in the ideal case. From Figure 6(b) and the green and violet curves representing the level of the first cascaded sidebands, one may think that their intensity is not significant at first sight when plotted on a linear scale and that no energy can build up into extra sidebands. However, when magnifying the low levels, one can see more clearly that those extra sidebands oscillate but remain typically below 2% of the total power of the central sidebands. In the case where we start from initial conditions where the three bands are in phase (see panel (d1) from Figure 6), the asymmetry parameter decreases steadily, and the evolution can be adjusted by an exponential trend. The asymmetry value thus tends asymptotically towards zero and then does not change sign. After a propagation distance of 6.60, the level of the two lateral sidebands becomes nearly equal (i.e. with a relative difference below 0.001).

The trends are slightly different when we consider the initial value $\phi_0 = \pi$. In that case, the asymmetry always decreases over the long term, but the evolution is not monotonic. Stages of decreasing asymmetry alternate with phases of increasing asymmetry, without however recovering its initial level. If we adjust the slow decay, we find a decay rate that is significantly lower than the rate obtained for components initially in phase (a factor of 17.60 is obtained between the two decrease rates). Additional numerical simulations carried out over much more significant distances (up to 64, results not shown here) have stressed the oscillatory behavior is only a transient stage and that for asymptotic propagation, the asymmetry value ultimately decreases monotonically down to 0.

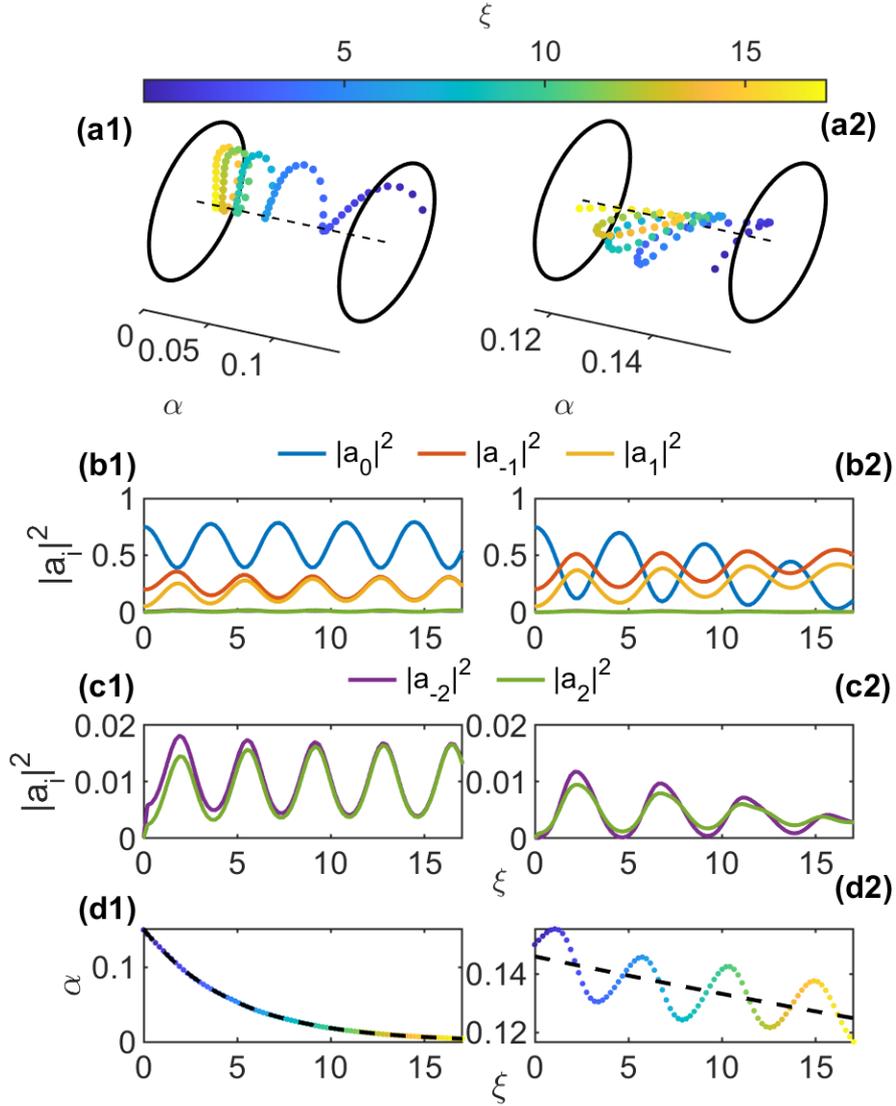

**Figure 6.** Results in the case of a propagation ruled by the NLSE. Results obtained for an initial phase of 0 and $\pi$ are displayed on panels **(1)** and **(2)**, respectively. **(a)** Trajectory plotted in the cylindrical system ($\eta$, $\phi$, $\alpha$). The distance of propagation is highlighted by changing the color of the symbols (see colormap). **(b)** Evolution of normalized intensity of the five inner spectral bands plotted on a linear scale. **(c)** Evolution of normalized intensity of the five inner spectral bands but plotted on a logarithmic scale. **(d)** Evolution of the asymmetry parameter (the colors refer to the distance as used in the colormap of (a)). Dashed lines refer to the fit of an exponential decrease.

To take our physical understanding of the destabilization of asymmetry a step further, we turn to a more complete model that takes into account the existence of $a_{\pm 2}$ components, even if they only covey a very small amount of the total energy. Thus, the NLSE can be rewritten by considering, instead of a field given by Equation (3), a field containing 5 spectral components:

$$A(\xi,\tau) = \sum_{n=-2}^{2} A_n(\xi)\exp(-in\Omega\tau). \qquad (9),$$

leading to the set of coupled equations describing the evolution of the various components : [22, 26, 27]

$$-i\frac{dA_j}{d\xi} + \frac{1}{2}(j\Omega)^2 A_j = \left(|A_j|^2 + 2\sum_{\substack{k\neq j \\ k=-2}}^{2}|A_k|^2\right)A_j \\ + \sum_{l,m,n}{}^*d_{lmn}A_l A_m A_n^* \qquad (10)$$

where $j,l,m,n = 0, \pm1, \pm2$ and $l,m \neq n$. Here $\sum_{l,m,n}^{*}$ denotes the permutations of the indices $l$, $m$, and $n$ such that $j = l+m-n$. The quantity $d_{lmn}$ is a degeneracy factor that is unity when $l=m$ and 2 when $l \neq m$.

In this framework, new interactions between the various components occur compared to the ones described in Figure 1(b). This is a key difference point compared to Equation (4), that is even if the energy in the extra sidebands is almost negligible, their existence allow new paths of nonlinear mixing that result eventually in different dynamics. By taking into account the interactions (a) and (b) described in **Figure 7** for the sidebands $a_{-2}$ and $a_{2}$, respectively, and using the previously described normalization process, one may obtain the following differential equations to describe the evolution of the sidebands $a_{\pm2}$:

$$\frac{\partial \tilde{a}_{\pm2}}{\partial \xi} = i\sqrt{\eta}\left[\left(\frac{1-\eta+\alpha}{2}\right)e^{i\phi} + \sqrt{(1-\eta)^2 - \alpha^2}\right] \\ \times \exp(-2i(1\pm\kappa)\xi)\,\exp(i(\varphi_0 + \varphi_{\pm1} - \varphi_{\mp1})) \qquad (11)$$

where the variables are written in a rotating frame $\tilde{a} = a \exp(-2i(1+\kappa)\xi + i\varphi_2)$ which results in simplification of terms $\frac{1}{2}(j\Omega)^2 A_j$ and $2\sum_{\substack{k \neq j \\ k=-2}}^{2} |A_k|^2 A_j$, and the appearance of the respective phase factor. Note that we consider an initial spectral amplitude of $a_{\pm 2}$ equal to zero. In Equation (11), we have not included the interactions sketched in panel (c) of Figure 7. Indeed, as the spacing between $a_{\pm 2}$ components and the pump is $2\Omega$, the energy transfer is inefficient (for $\kappa = -1.2$, the $a_{\pm 2}$ components lie above the cutoff frequency of the modulation instability gain bandwidth). We also neglected the interactions that do not involve the pump (interaction between $a_{\pm 1}$ and $a_{\pm 2}$).

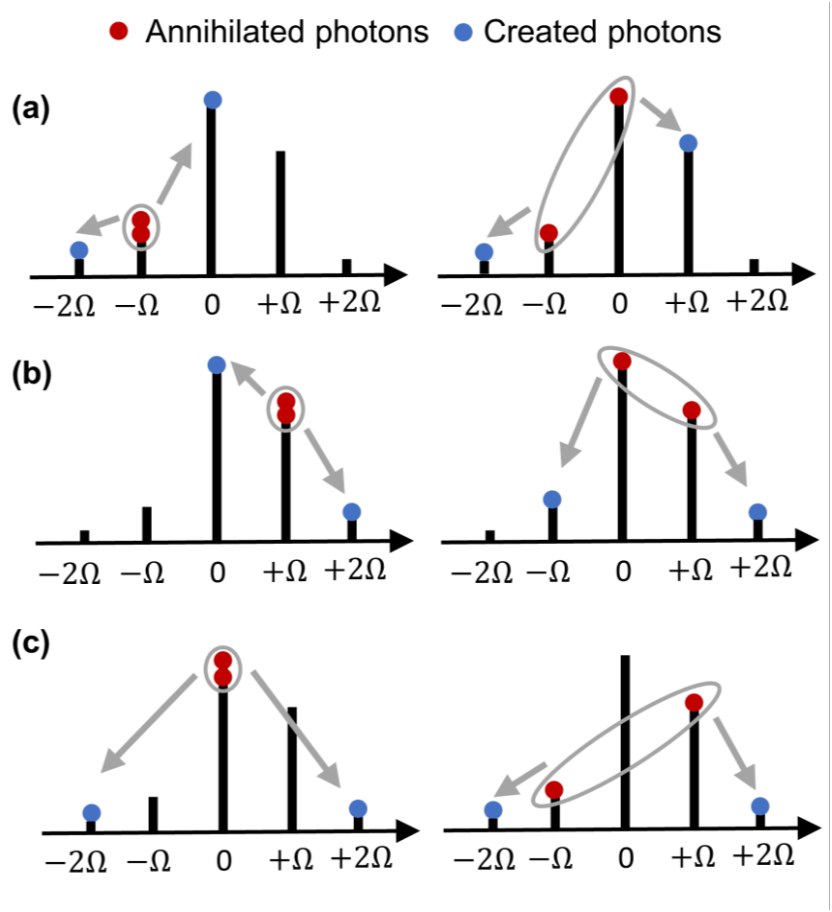

**Figure 7.** Various new combinations that develop when five spectral components are taken into account.

As the propagation distance $\Delta\xi$ over one iteration is short enough, the growth of the higher-order sidebands $a_{\pm 2}$ does not impact significantly the level of the other involved parameters so that the later can be considered as fixed to a first approximation. Also the resulting phase accumulation in $\exp(-2i(1+\kappa)\Delta\xi)$ and we can consider a starting point with $\varphi_{-1}=\varphi_1$ and $\varphi_0=0$ so phase difference can be simplified. In that context, one may approximate the differential operator by a finite difference. Moreover, as $\Delta a_{\pm 2} \simeq |\Delta \tilde{a}_{\pm 2}|$, one may obtain a fully explicit equation for the level $|\Delta a_{\pm 2}|^2$ reached by the sideband after propagation in the segment of length $\Delta\xi$:

$$|\Delta a_{\pm 2}|^2 = \Delta\xi^2\, \eta_0 \frac{1-\eta_0 \mp \alpha_0}{2}\left[\frac{5-5\eta_0 \pm \alpha_0}{2} + 2\cos(\phi_0)\sqrt{(1-\eta_0)^2 - \alpha_0^2}\right]. \quad (12)$$

From these results illustrated in **Figure 8**, panel (a) for an initial asymmetry of $\alpha_0=0.15$, it is clear that the initial point $(\eta_0, \phi_0)$ will affect the strength of the growth of the additional sideband and that the right part of the phase portrait will be much more affected than the left counterpart by the growth of the additional sidebands. Comparing the results of Equation (12) with the predictions obtained from the numerical integration of the NLSE over one iteration distance (panels b), one may note the excellent qualitative agreement between the approximate analytical model and the results taking into account all the possible nonlinear interactions without any simplifying assumptions. Note that given the condition linking $\eta_0$ and $\alpha_0$, the phase plane cannot be explored for values $\eta_0>0.85$.

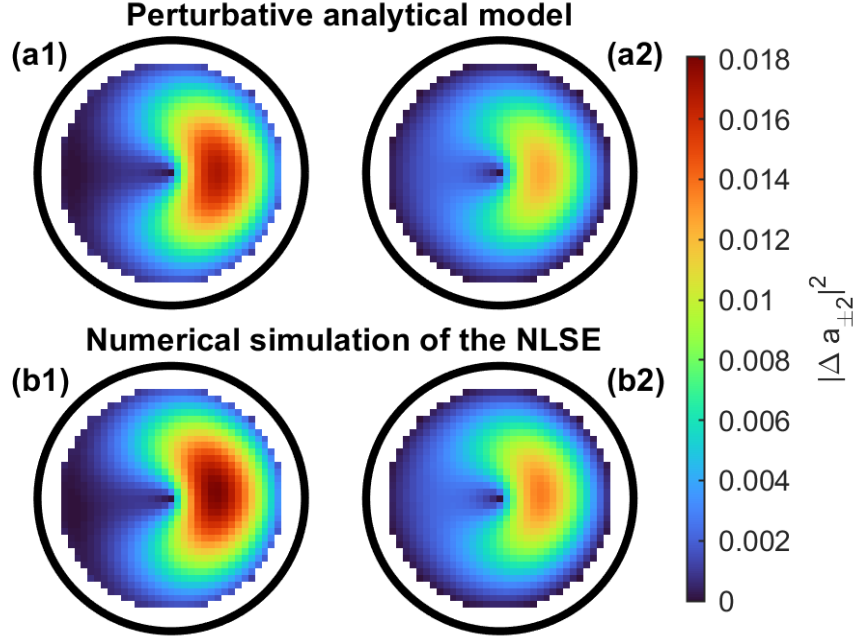

**Figure 8.** Growth of the spectral sidebands $|\Delta a_{\pm 2}|^2$ (panels 1 and 2, respectively) after a propagation distance of $\Delta \xi = 0.21$ according to the starting value on the phase portrait ($\eta_0$, $\phi_0$) for an initial value of the asymmetry $\alpha_0 = 0.15$. Results of the perturbative analytical model (Equation (12)) are compared with the results obtained from the numerical integration of the NLSE.

We also derived the evolution of the asymmetry from the set of coupled Equations (10). Stating that

$$\frac{d\alpha}{d\xi} = \frac{d|a_{-1}|^2}{d\xi} - \frac{d|a_{+1}|^2}{d\xi}$$
$$= a_{-1}^* \frac{da_{-1}}{d\xi} + a_{-1} \frac{da_{-1}^*}{d\xi} - a_{+1}^* \frac{da_{+1}}{d\xi} - a_{+1} \frac{da_{+1}^*}{d\xi}, \quad (13)$$

one may obtain the following equation:

$$\begin{aligned}\frac{d\alpha}{d\xi} = &- 4\, a_{-1}^2\, a_{-2} a_0 \sin(\varphi_0 + \varphi_{-2} - 2\varphi_{-1}) \\ &- 4\, a_{-2}\, a_1 a_{-1} a_0 \sin(\varphi_{-2} + \varphi_1 - \varphi_{-1} - \varphi_0), \\ &+ 4\, a_1^2\, a_2 a_0 \sin(\varphi_0 + \varphi_2 - 2\varphi_1) \\ &+ 4\, a_2\, a_1 a_{-1} a_0 \sin(\varphi_2 + \varphi_{-1} - \varphi_1 - \varphi_0)\end{aligned} \quad (14)$$

It is then clear that the presence of the higher-order sidebands is required to observe a non-zero derivative of $\alpha$. As the level of the higher-order sidebands remains well below -17dB, a perturbative approach can once again be implemented. Plugging the approximate solution of Equation (12) into (14) (here we consider for to the value of $a_{\pm 2}$ its average value $\Delta a_{\pm 2}/2$ over the segment), one obtains using normalized notations :

$$\Delta\alpha = \frac{\eta_0 \Delta\xi^2}{2}\left[\left(1-\eta_0-\alpha_0\right)^{3/2}\left(\sqrt{1-\eta_0-\alpha_0}+3\sqrt{1-\eta_0+\alpha_0}\cos\phi_0\right)\right.\\ \left.-\left(1-\eta_0+\alpha_0\right)^{3/2}\left(\sqrt{1-\eta_0+\alpha_0}+3\sqrt{1-\eta_0-\alpha_0}\cos\phi_0\right)\right], \quad (15)$$

The results are plotted in panel (a) of **Figure 9** and show good qualitative agreement with the results computed using the exact numerical integration of the NLSE over one segment of fiber. The remaining discrepancies that exist are attributed to the perturbative approach that we involve both to obtain $\Delta a_{\pm 2}$ as well as $\Delta\alpha$ and to the fact that the interactions (c) (see Figure 7) are not taken into account in the derivation of Equation (11). The results highlight that both sides of the separatrix exhibit very different behavior. First of all, when operating on the right side, values of $\Delta\alpha$ are always negative so that one may expect a trajectory occurring on that side to experience a monotonic evolution as exemplified by the trajectory plotted on panel (b), (solid black line), leading to the monotonic evolution plotted on panel (d1) of Figure 6. This contrasts radically with the behavior observed when operating on the left side where the trajectory (see dashed black line on panel b) experiences alternatively $\Delta\alpha$ with positive and negative values, leading to the oscillatory evolution plotted in Figure 6, panel (d2). One explanation of these very different features lies in the fact that for the right part of the phase plot, the phase remains bounded so that the cosine term involved in Equation (15) is always positive. On the contrary, on the left side, the phase is unbounded and the cosine term therefore explores both positive and negative values. From Equation (15), one can also make out that an opposite value of asymmetry $\alpha_0$ will lead to opposite values of $\Delta\alpha$. When the initial asymmetry becomes null, $\Delta\alpha$ also vanishes, which is consistent with the observations of Figure 6(d), $\alpha = 0$ being a stable configuration.

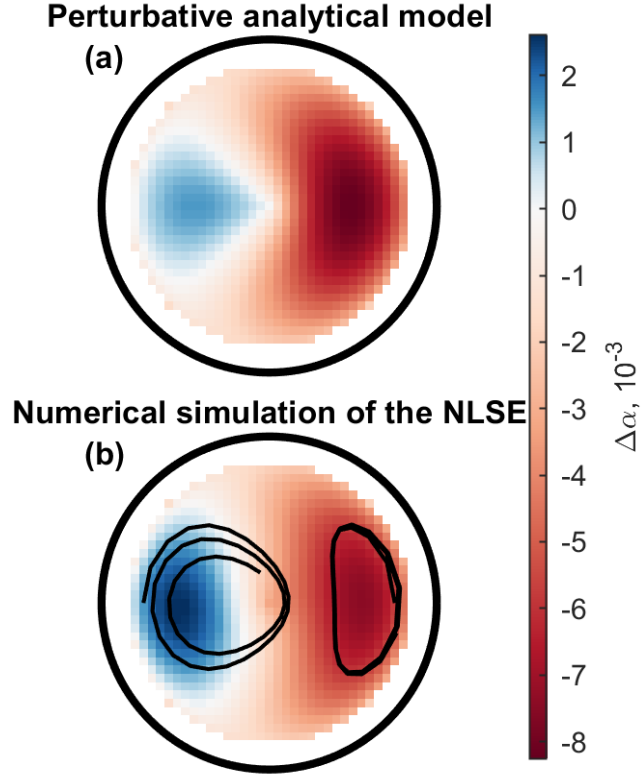

**Figure 9.** Change in the asymmetry parameter over the phase plane for an initial asymmetry of 0.15. The approximate analytical solution (Equation (15), panel **(a)**) is compared with the results obtained from the numerical integration of the NLSE (panel **(b)**).

## 4. Experimental setup and results

### 4.1. Setup and experimental limitations

Experimentally recording the longitudinal evolution of the optical field is a challenging task. Various methods have been implemented in the case of the analysis of the dynamics occurring in the full NLSE system with symmetrical input conditions: destructive cut-back measurements,[28] the involvement of multiple fiber segments,[29] distributed optical time domain reflectometry [30, 31] or evolution in a recirculating loop.[32, 33] To the best of our knowledge, none of these previous experimental studies has investigated the impact of asymmetric conditions. In all cases and even when dealing with symmetric conditions, management of optical losses is a critical issue and requires additional distributed gain in order not to deviate from the loss-free hypothesis.[30] Moreover, the discrepancy between experiments and ideal FWM dynamics becomes significant very quickly with a rise of the additional marked sidebands resulting from unwanted cascaded FWM.

The experimental setup we developed to test the concept presented in 2.2 is shown in **Figure 10** and is made of commercially available telecommunications components. First, a laser operating at 1550 nm emits a continuous wave (CW) of high coherence required to observe optimal four-wave mixing.[34] A phase modulator driven by a 40 GHz RF sinusoidal modulation converts the monochromatic laser spectrum into a set of equally spaced coherent spectral lines.[35] The resulting symmetrical comb is then processed using a programmable filter (Finisar/II-VI Waveshaper device based on liquid crystal on silicon [36]) that carries out several operations: elimination of unwanted spectral components, the precise adjustment of the ratio $\eta_n$ between the central and lateral components as well as imprinting the targeted asymmetry $\alpha_n$. The device also simultaneously implements the relative phase $\phi_n$. The tailored three-component signal with the target ($\eta_n$, $\phi_n$, $\alpha_n$) is then amplified by an erbium-doped fiber amplifier that delivers a tunable average power that does not depend on the input spectral content (and thus enables us to reach various $\kappa$ values, typically between -2.51 and -1.0).

Nonlinear propagation takes place in single-mode optical fiber with dispersion and nonlinear parameters being respectively -7.6 ps$^2$.m$^{-1}$ and 1.7 W$^{-1}$.km$^{-1}$. The fiber length is 500 m, with this length selected as a tradeoff between the sensitivity of the detection stage of our setup and the appearance of higher order nonlinear effects, such as Brillouin scattering.

The output signal is then attenuated and split into two in order to record both its spectral phase and intensity. An optical spectrum analyzer (OSA, resolution 0.1 nm) provides directly the ratio $\eta_{n+1}$ and $\alpha_{n+1}$. The spectral phase offset $\phi_{n+1}$ is retrieved from the temporal delay between the central and lateral sidebands as measured with a high-speed sampling oscilloscope. The experimentally measured values is then imprinted as new input values and the process can be iterated at will without any accumulation of deleterious amplified spontaneous emission and without any significant growth of unwanted spectral sidebands or noise. We typically investigated propagation distances up to 50 km, corresponding to 100 iterations. Note that we recently proposed an alternate non-iterative scheme benefiting from machine learning strategies based on a large set of initial random measurements that are then processed using artificial neural networks.[19]

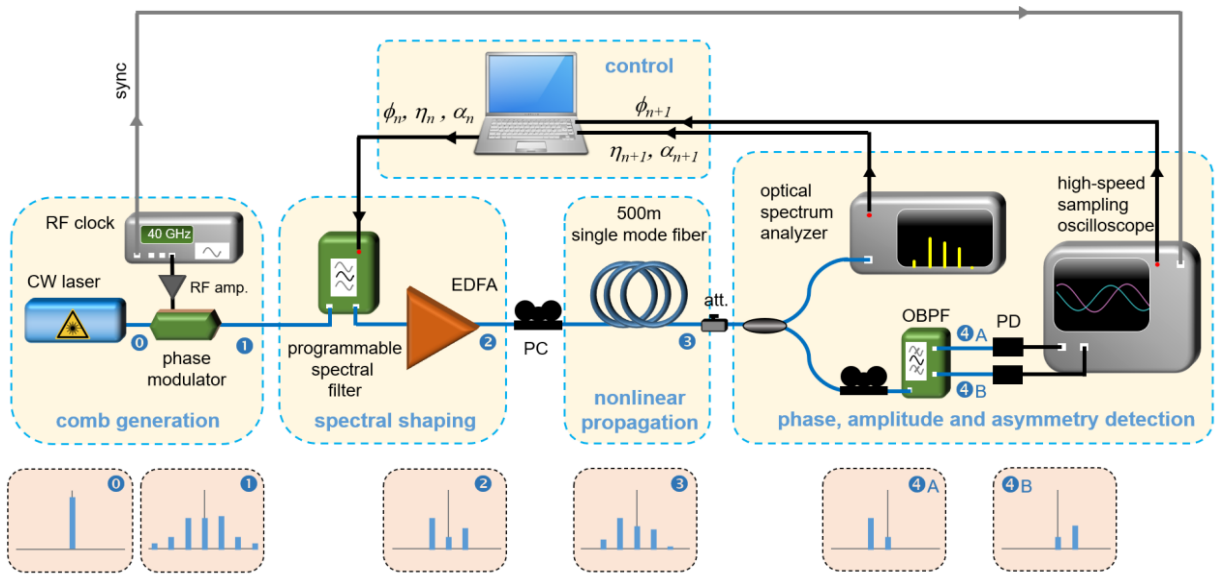

**Figure 10.** Implemented experimental setup. ASE: Amplified Spontaneous Emission source, PC: Polarization Controler, DCF: Dispersion Compensating Fiber, DSF: Dispersion-Shifted Fiber, VA: Variable Attenuator, PD: Photodiode, OSA: Optical Spectrum Analyzer

For the study we report here, one limitation of our experimental setup has to be kept in mind: the sensitivity of our spectral detection. We can estimate that we have any accuracy of the order of 0.01 on the value of $\alpha$. Even if this value seems very low at first sight, its accumulation over iterations may become a problem. This leads us to favor experimental conditions leading to a rapid decrease in the asymmetry. We have therefore focused our studies on a value of $\kappa = -1.2$ and for an initial condition that is already highly asymmetric ($\alpha_0 = 0.15$). Unfortunately, even in this case, the possible accumulation of small errors at each iteration prevents us from studying with confidence the left-hand side of the separatrix, where the decay rate is significantly lower than for the right-hand side. Therefore, in order to exclude measurements with moderate accuracy, we have restricted our study to the part to the right of the separatrix where the phase is bounded.

### 4.2. Experimental results

The experimental results recorded for $\eta_0 = 0.75$ and $\alpha_0 = 0.15$ are shown in **Figure 11** where we can see that the trends predicted by numerical simulations of the NLSE (see Figure 6(d1)) are convincingly reproduced. The level of the three main sidebands oscillates and the difference between the $a_{-1}$ and $a_{+1}$ components tends to decrease. The amplitude of the $a_{\pm 2}$ sidebands is negligible at first sight and is only visible when plotted on a magnified view (panel b). The asymmetry decreases with a speed close to the one expected from numerical simulations. The small discrepancies can be ascribed to the limited accuracy of the experimental detection. The 3D view of the trajectory is also in good qualitative agreement with the numerical expectations.

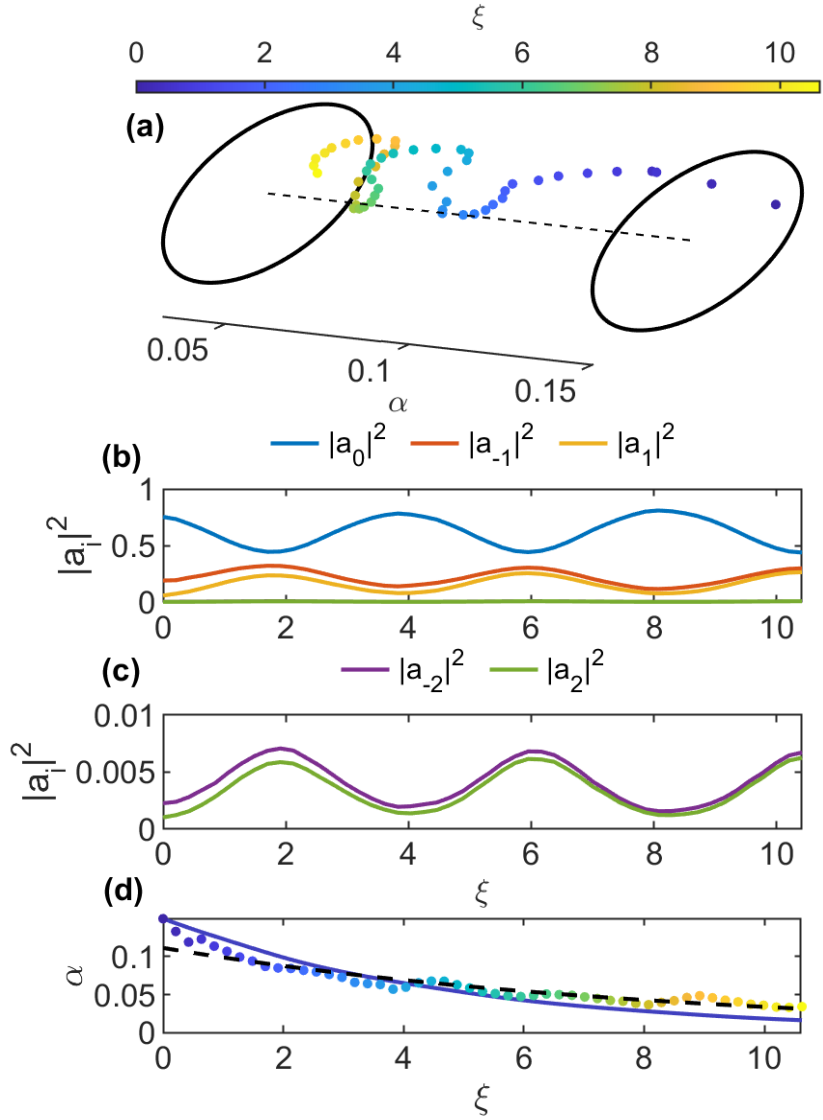

**Figure 11.** Experimental evolution of the system parameters: (a) Evolution of the parameters plotted in cylindrical coordinates (b) level of the normalized sidebands plotted on a linear scale (c) level of the normalized sidebands plotted on a logarithmic scale (c) Evolution of the asymmetry parameter. The experimental results are fitted with an exponential function (dashed line). Numerical simulations from Figure 6(d1) are shown with a full (blue) line.

Some additional measurements are shown in **Figure 12**. Carrying a large number of random measurements on the phase plane (500 points), we were able to measure the growth of the lateral sidebands $\Delta a_{\pm 2}$ (here, recorded for a null initial asymmetry). For improved visualization, the discrete set of measurements (panel (a1) is interpolated using the universal interpolation features of a simple neural network,[37] leading to panel (a2). The experiments clearly reproduced the qualitative expected trends recalled in panel (a3), with a much more pronounced growth of the extra-sidebands for the right side of the separatrix.

We also performed measurements for the longitudinal evolution of the asymmetry coefficient for various values of $\eta_0$ and for $\alpha_0 = \pm 0.15$, the initial sidebands being kept in phase. The experimental trends reported in panel (b1) of Figure 12 are in qualitative agreement with the one obtained numerically from the simulation of the full NLSE (panel (b2)). They confirm that an opposite value of the initial asymmetry leads to a mirror evolution, and that the absolute value of the asymmetry is also reduced over propagation, tending asymptotically to values close to zero. The results also stress that the decrease rate is not dramatically dependent on the value of $\eta_0$.

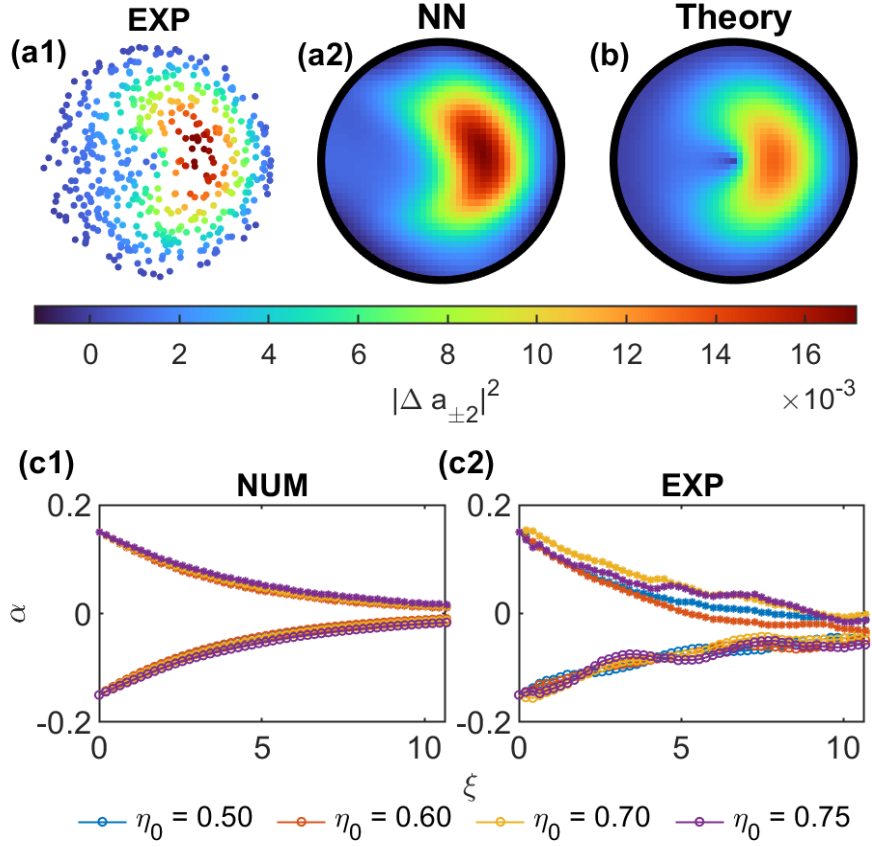

**Figure 12. (a)** Growth of the spectral sidebands $|\Delta a_{\pm 2}|^2$ (panels 1 and 2, respectively) after a propagation distance of $\Delta \xi = 0.21$ according to the starting value on the phase portrait ($\eta_0$, $\phi_0$) for an initially symmetric waveform. The experimental measurements **(a1)** are interpolated using a neural network **(a2)** and compared with the perturbative analytical model (**a3**, Equation (12)). **(b)** Evolution of the asymmetry parameter for initial values $\alpha_0 = \pm 0.15$ and for different values of $\eta_0$. Numerical simulations are compared with experimental results (panels **c1** and **c2**, respectively).

## 5. Conclusion

In conclusion, we have investigated in this paper how the asymmetry parameter between the three principal spectral components evolves in systems involving the four-wave mixing process. Whereas this quantity is conserved in an ideal system without any additional cascaded mixing, the generation of spectral components in the framework of the non-linear Schrödinger equation leads to different and more complex dynamics where the asymmetry fluctuates widely, possibly changing sign. The study of a third system based on iterated propagation and much closer to the ideal case revealed that the idealized description misses some key interaction paths that ultimately result in the loss of the asymmetry, even if the extra sidebands are kept unsignificant. Indeed, even a very weak generation of additional spectral components is enough to affect the asymmetry conservation, leading to a decay of this parameter down to a zero value and stabilizing the Fermi-Pasta-Ulam-Tsingou recurrent behavior. A perturbative analytical approach enabled us to gain a physical understanding of the physical processes that break this conservation law, as well as the impact of the initial phase. The dependence of the observed behavior on the position in the phase plane was also studied and these various trends were confirmed experimentally. Therefore, the modification of the difference level between the first lateral sidebands reveals existence of cascaded four-wave interactions.

The study highlights the great care that has to be taken when discussing the conservation of invariants. While the growth of sidebands may seem insignificant at first sight, given their 50-fold smaller amplitude, their cumulative effect nevertheless has a very real impact that cannot be overlooked. We have emphasized their impact on the asymmetry parameter but one can also mention other strong consequences such as the impossibility to generate a stationary wave in the system (at least not until the asymmetry has fully dropped down to zero). Indeed, the fixed point existing in the ideal four-wave mixing [12] is no longer fixed due to progressive drop of the asymmetry down to zero. Other open questions remain such as the modelling of the dynamics over an arbitrary large distance by a single set of differential equations rather than the use of approximate solutions only valid over a short propagation distance. In that context, involving data driven discovery using sparse regression may provide some powerful clues.[38] For the typical range of parameters we have here considered ($-4 < \kappa < -1$), potential cascaded sidebands do not fall within the modulation instability gain bandwidth of the pump. However, the picture may become much more richer when considering $-1 < \kappa < 0$ and require a non-perturbative treatment.[25] Our study was primarily focused on degenerate four-wave mixing but the approach can be straightforwardly extended to the non-degenerate case.[26, 39] Our conclusions can also be directly of interest to other physical systems implying four-wave mixing interactions and reduced models, as it has been developed in the field of hydrodynamics for example.[40, 41] We also anticipate it can be helpful in the framework of three wave interactions and quadratic nonlinearities.[42, 43]

**Acknowledgements**

We thank Bertrand Kibler for fruitful discussions during the initial stages of the project. The work was funded by the Agence Nationale de la Recherche (Optimal project - ANR-20-CE30-0004; I-SITE BFC - ANR-15-IDEX-0003), the Région Bourgogne-Franche-Comté, the Institut Universitaire de France and the Centre National de la Recherche Scientifique (MITI interdisciplinary programs, 'Evenements extrêmes').